\documentstyle[aps,prl,twocolumn]{revtex}

\include{psfig}  
\begin{document}
\draft \title{\bf Dynamics in a supercooled molecular liquid: 
Theory and Simulations}
\author{A. Rinaldi, F. Sciortino and P. Tartaglia}
\address{Dipartimento di Fisica and Istituto Nazionale
 per la Fisica della Materia, Universit\'a di Roma {\it La Sapienza},\\
 P.le Aldo Moro 2, I-00185, Roma, Italy} 
\date{\today} \maketitle

\begin{abstract}
We report extensive simulations of liquid supercooled states for 
a simple three-sites molecular model, introduced by Lewis and Wahnstr\"om 
[L. J. Lewis and  G. Wahnstr\"om, Phys. Rev. E {\bf 50}, 3865 (1994)] 
to mimic the behavior of {\it ortho}-terphenyl. The
large system size and the long simulation length allow 
to calculate very precisely --- in a large 
$q-$vector range --- self and collective correlation functions,
providing a clean and simple reference model for
theoretical descriptions of molecular liquids in
supercooled states. The time and wavevector dependence of 
the site-site correlation functions are compared with detailed
predictions based on ideal mode-coupling theory,  
neglecting the molecular constraints. Except for the 
wavevector region where the dynamics is controlled by the center
of mass (around $9$ $nm^{-1}$), 
the theoretical predictions compare very well with the
simulation data. 
\end{abstract}

\section{Introduction}

In the last decade, increased computational resources have
been used to tackle the study of the onset of glassy 
dynamics\cite{KobAnd,Kob-LJ,self,collective,madden,heuer-doliwa,lai,kob-dumb}, 
one of the most interesting open problems in the
physics of liquids. Supercooled liquids are indeed characterized by 
an extreme $T$-dependence of the structural times, which covers
more than 15 orders of magnitude in a small temperature interval.
Present computer facilities allow 
to follow the change in the structure and in the
dynamics of the system (in equilibrium) over a $T$ range where the
diffusivity changes more than 5 order of magnitudes. 
At the same time, computer simulations 
are starting to provide a detailed 
picture of the structure of the
potential energy landscape probed in supercooled 
states\cite{heuer,sri,skt,claudio,lanave,selle}.
The calculated trajectories offer an
outstanding possibility for studying the origin of the slowing down
of the dynamics, and call for a careful comparison between theories
and ``exact'' numerical calculations. 

``Exact'' results calculated for model-system
and theoretical  predictions based on 
mode coupling theory (MCT)\cite{review-glass,goetze-pisa} 
have been reported for several atomic liquids, 
---  encompassing hard spheres, soft
spheres, Lennard-Jones\cite{Kob-LJ}, and recently silica\cite{SIO2}.
In the case of  molecular liquids, 
where the possibility of comparing with experiments makes
the effort even more valuable, molecular dynamics (MD) 
simulations extending to the
100 $ns$ timescale are now possible. Detailed comparisons between
theories and simulations appears as 
well\cite{ssm,mmct2,theis-new,winkler}.

MD simulations have often been used to generate
reference systems for validating theories of liquids, starting from
the pioneering work of Alder and Wainwright\cite{alder} on hard spheres.
A very interesting model for a molecular liquid was designed by
Lewis and Wahnstr\"om  (LW)\cite{lw}
by  gluing in a rigid molecule three identical Lennard Jones (LJ)
atoms. The shape of the molecule (an isoscele triangle) and the
LJ parameters were chosen to mimic as close as possible 
one of the most studied 
glass-forming, 
liquid\cite{plazek94,sillescu91,sillescu93,sillescu95,toelle97,masciovecchioPRL99,monacoPRL98,monacoPRL99}
{\it ortho}-terphenyl (OTP).
This model provides a clean bridge between studies based on
atomic LJ potentials and more complex molecular potentials.
From a theoretical point of view, such a model is also very valuable,
since the LW model lends itself to several approximation of
increasing complexity. For example, one could consider
the LW system as a liquid of identical atoms, correlated in space 
according to the site-site correlation function (without
differentiating between intra and intermolecular sites), or one can
improve by including the
molecular correlation via the triplet site-site-site correlations. 
Solution of the exact molecular model, implementing a site-site
description\cite{site-MCT} which accounts for the intramolecular constraints 
or an expansion in generalized spherical 
harmonics\cite{franosch-dumb,schilling}
is also foreseeable.

The LW potential was introduced in 
Ref.\cite{lw}, which
reports a  study of the slow dynamics of the LW liquid. The major properties
of such models were discussed both for self and collective
properties. In following papers\cite{walstrom2}, 
LW made also some preliminary attempts to  
describe the intermediate time dynamics using MCT 
predictions. Favored by the increased computational power, 
we have decided to revisit such a model and --- by extending the
simulation time by a factor  20 and the number of molecules
by a factor 30 --- to calculate detailed properties
of the dynamics in supercooled states, like the non ergodicity
parameters for both self, collective and orientational properties
to be used as reference system to be compared with theoretical
predictions.  We present such data in this article. We also
solve the MCT equations for such a model
using as input the site-site structure factor  and the
site-site-site triplet correlation functions calculated from
the simulated trajectories. 
The comparison between the molecular dynamics data and the MCT
predictions is very valuable in assessing the
role of the center of mass dynamics (which is neglected in
the chosen MCT approach), the role of the
triplet correlations\cite{barrat-c3,SIO2} and, in general, the
ability of ideal MCT to capture the dynamics of molecular
liquids in weakly supercooled states.

\section{The model and numerics}

The geometry of the LW molecule is a rigid 
three sites isoscele triangle hosting 
on each site a Lennard Jones atom. 
The site-site spherical potentials is

\begin{equation}
V(r)= 4 \epsilon\,( ({{\sigma}\over {r}})^{12} -({{\sigma}\over {r}})^6)
+ A + B\, r 
\end{equation}

\noindent
with $\epsilon=5.276$ $kJ/mol$, $\sigma= 0.4828$ $nm$,
$A=0.4612$ $kJ/mol$ and  $B= -0.3132$ $kJ/mol/nm$.
The parameters of the potential have been selected to reproduce
bulk properties of the OTP molecule\cite{lw}. The value of $A$ and  $B$
has been selected to bring the potential and its first derivative to
zero at $r=1.2616$ $nm$. The resulting potential has a minimum at
$r=0.5421$ $nm$ of depth $-4.985$ $kJ/mol$.
The length of the two short sides is $0.483$ $nm$ and of
the long side $0.588$ $nm$, forming an isoscele angle of
$75$ degrees. 

We have studied a system composed by $N=9261$ molecules 
for several state points, from $T=255$ $K$ to $T=346$ $K$, 
as listed in Table \ref{table:table}. 
The choice of a large system size was performed to avoid  spurious
oscillations in the correlation functions introduced by the periodic
boundary conditions. 
The large size was also preferred to
access information --- here and in future analysis --- 
on the small wavevector dynamical behavior.

In this article we discuss in detail the auto correlation function
for the collective and self density operators, 
both for center of mass (COM) and sites (s). The density operators
$\rho$
(and their corresponding autocorrelation functions) are 
defined as\cite{hansen}
\begin{eqnarray}
\rho_{_{COM}}({\bf q}) \equiv {{1}\over{\sqrt{N}}}  
\sum_{i=1}^{N} e^{i\, {\bf q}\cdot {\bf r}^i_{_{COM}}};\nonumber
\\
 F_{_{COM}}({\bf q},t)\equiv  
\langle \rho_{_{COM}}({\bf q},t) \rho_{_{COM}}({\bf q},0)^* \rangle
\end{eqnarray}
\begin{equation}
\rho_s({\bf q})\equiv{{1}\over{\sqrt{3N}}} \sum_{i=1}^{N} \sum_{j=1}^3 e^{i\, 
{\bf q}\cdot {\bf r}^i_j};~~ F_{s}({\bf q},t)\equiv  
\langle \rho_{s}({\bf q},t) \rho_{s}({\bf q},0)^* \rangle
\end{equation}

\begin{eqnarray}
\rho_{_{COM}}^{self}({\bf q})\equiv e^{i\, {\bf q}\cdot {\bf r}^i_{_{COM}}};\nonumber
\\
~~F_{_{COM}}^{self}({\bf q},t)\equiv  
\langle \rho_{_{COM}}^{self}({\bf q},t) \rho_{_{COM}}^{self}({\bf q},0)^* \rangle
\end{eqnarray}
\begin{equation}
\rho_s^{self}({\bf q})\equiv e^{i \,{\bf q}\cdot {\bf r}^i_j};~~F_{s}^{self}({\bf q},t)\equiv  
\langle \rho_{s}^{self}({\bf q},t) \rho_{s}^{self}({\bf q},0)^* \rangle
\end{equation}

\noindent
where ${\bf r}^i_{_{COM}}$ and $ {\bf r}^i_j$ are the position of the COM and
of the $j$-site of molecule $i$.  Moreover, $S_{_{COM}}(q)\equiv F_{_{COM}}(q,0)$
and  $S_{s}(q)\equiv F_{s}(q,0)$ are the COM and site static structure factors.

We also present the $l$-dependent rotational 
correlators for the molecule symmetry axis $\bf {\mu}$, defined as
\begin{equation}
C_l(t)\equiv\langle P_l[cos\theta(t)]\rangle
\end{equation}
where  $\theta(t)=cos^{-1}
[\langle {\bf \mu} (t) \cdot {\bf \mu}(0)\rangle/\langle{\bf \mu}(0) \cdot 
{\bf \mu}(0)\rangle]$ and $P_l$ is the $l$-order Legendre polynomial.
The triplet site-site-site 
correlation function $c_3(\bf k,\bf p \bf q)$ 
defined as 
\begin{eqnarray}
\langle \rho_s({\bf k}) \rho_s({\bf p}) \rho_s({\bf q}) \rangle \nonumber 
\\
= 3 N S_s(k) S_s(p) S_s(q) \delta_{{\bf q,k+p}} 
(1 + n^2 c_3({\bf k},{\bf p} {\bf q}))
\label{eq:c3}
\end{eqnarray}
has been calculated evaluating the left hand side of Eq.~\ref{eq:c3};
$n$ is the site number density. The evaluation of 
$c_3$ has been performed at $T=266$ $K$ 
averaging over 1000 independent configurations, spanning a time interval of
75 $ns$.  Where possible, up to 300 different
triplets of wavevectors with the same moduli and relative
angles have been averaged. We have calculated $c_3$ for  moduli 
of $\bf k$,$\bf p$ and $\bf q$ less than $44.4$ $nm^{-1}$
with a mesh of $0.44$ $nm^{-1}$.

\section{Mode Coupling Approximations}
\label{sec:mct}
In this article we present two theoretical calculations
of the dynamics of the sites.
The first approximation neglects completely the
presence of the molecular constraints and 
assumes that the site-site structure factor includes all
requested information on the structure of the liquid. 
This strong approximation allows to apply in a straightforward 
manner the MCT equation for simple liquids. According to MCT 
the time evolution of the normalized 
collective density correlation functions $\Phi_q(t)$ is
given by

\begin{equation} 
\ddot{\Phi}_q(t)+\nu_q \dot{\Phi}_q(t)+\Omega_q^2 \Phi_q(t)+
\Omega_q^2 \int_0^t ds\,m_q(t-s) \dot{\Phi}_q(s)=0.
\label{eq:dyn}
\end{equation}

\noindent
Here $\Omega_q = \sqrt{q v /S_s(q)}$, with $v$ 
denoting the thermal velocity,
is an effective phonon-dispersion law and $\nu_q=\nu_1 q^2$ denotes a damping constant. 
The kernel $m_q$ is given as 
$m_q(t) = {\cal F}_q (\{\Phi_k(t)\})$, where the mode-coupling
functional ${\cal F}_q$ is determined by the structure factor:
\begin{equation}
{\cal F}_q (\{f_k\}) = {1 \over 2} \int{ {{d^3k} \over 
(2 \pi )^3} V_{{\bf q},{\bf k}} f_k f_{|{\bf q}-{\bf k}|}},
\label{eq:mq}
\end{equation}
\begin{eqnarray}
V_{{\bf q},{\bf k}} \equiv  S_s(q) S_s(k) S_s({|{\bf q}-{\bf k}|}) {n \over {q^4}}
\left[ {\bf q} \cdot 
{\bf k}\,{c_k} +{\bf q} \cdot  
({\bf q}-{\bf k})\,{{c_{|{\bf q}-{\bf k}|}} }
 \right]^2.
\label{eq:v}
\end{eqnarray}
\noindent
where $c_q = (1-S_s(q)^{-1})/n$ is the  direct 
correlation function in $q$-space.

The second approximation is devised to retain some of the
information of the molecular shape through the inclusion of
the triplet correlation function $c_3$\cite{barrat-c3}, 
previously assumed to be zero. 
The memory function is then given by

\begin{eqnarray}
V_{{\bf q},{\bf k}} \equiv  S_s(q) S_s(k) S_s({|{\bf q}-{\bf k}|}) 
{n \over {q^4}}\nonumber
\\
\left[ {{\bf q}} \cdot 
{\bf k}\,{c_k} +{\bf q} \cdot  
({\bf q}-{\bf k})\,{{c_{|{\bf q}-{\bf k}|}} }
-q^2 n\, c_3({\bf q},{\bf k},{\bf q}-{\bf k})
 \right]^2.
\label{eq:vc3}
\end{eqnarray}
\noindent

For simple atomic liquids, previous studies\cite{barrat-c3} have shown
that the approximation $c_3=0$ does not modify the MCT predictions,
except for a small shift in the estimate of the critical
temperature. Recently, it has been shown\cite{SIO2} that in network
forming liquids, the $c_3$ contributions play a crucial role. 
Including the $c_3$ contributions could, in the present case,
be relevant for describing the intramolecular site-site relations.
For this reason, we compare the numerical data with both
predictions.

We numerically solved Eq.~\ref{eq:dyn} on a grid of 300 
equally spaced $q$ values extending up to $q=120$ $nm^{-1}$, implementing
the efficient techniques described in Ref.\cite{numerics}.
The long time limit of  $\Phi_q(t)$  at the dynamical critical temperature ---
the so-called non ergodicity parameter -- 
$f_q$ is obtained by an iterative solution of the bifurcation equation
\begin{equation}
{{f_q} \over {1-f_q}}={\cal F}_q(\{f_k\}).
\label{eq:fq}
\end{equation}
Close to the critical point, 20000 iterations where requested to
solve Eq.\ref{eq:fq} with a precision of $10^{-15}$.
We have solved Eqs.\ref{eq:dyn} for both approximations. The
estimated critical temperatures are $T=152$ $K$ and 
$T=340$ $K$ for the two approximations. Thus, neglecting completely the
molecular constraints (first approximation) strongly weakens the
vertex function, resulting in a theoretical critical temperature
lower than the numerical $T_{MD}^c$. Including the triplet correlation
moves up the critical temperature by a factor bigger than 2, resulting in
an overestimation of the critical temperature, as always found 
in all previous MCT calculations. 
Notwithstanding the large difference in critical temperature
the exponent parameter\cite{review-glass} is $\lambda=0.70$ 
in both approximations,
corresponding to the exponents $b=0.64$ and $a=0.32$\cite{review-glass}.

\section{Molecular Dynamics Results}

\subsection{Static}

\begin{figure}
\centerline{\psfig{figure=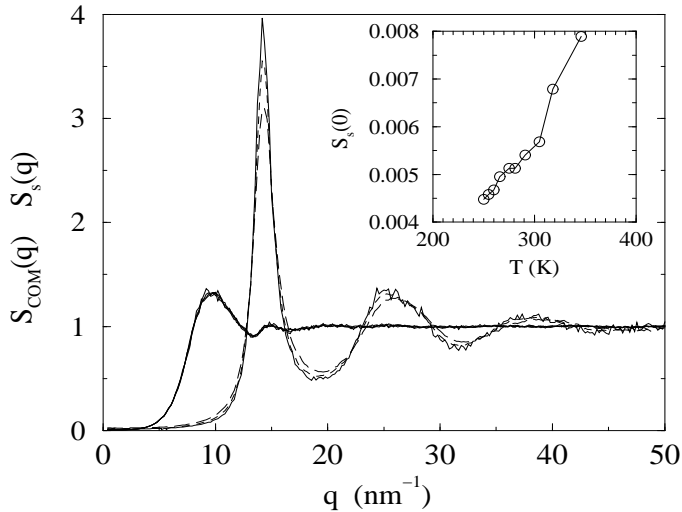,height=7.cm,width=9.cm,clip=,angle=0.}}
\caption{$S_s(q)$ and $S_{_{COM}}(q)$. For clarity reasons, only
data for $T=255 K$, $T=281 K$ and $T=346 K$ are shown.  
Note that $S_{_{COM}}(q)$ is
temperature independent, while for the  $S_{s}(q)$, the height of the
first peak increases on cooling. The inset shows the  $T$-dependence of the
structure factor at $q=0$.
}
\label{fig:sq}
\end{figure}

The site and COM static structure factors are shown in
Fig. \ref{fig:sq}, as a function of temperature.  The COM structure
factor does not show any appreciable change with $T$. Instead, the
site structure factor increases significantly at the first peak
position (corresponding in real space to the site-site nearest
neighbor distance), consistently with the behavior of simple liquids.
The absence of any change in the COM structure factor shows that the
slowing down of the dynamics in this model is controlled by the
changes in orientational order. It is the ordering process in the
orientational degrees of freedom which drives the slowing 
down of the dynamics. Any theoretical approach attempting to predict
the slow-dynamics in this model must require orientational
(or site-site) static  information as input.
The absence of any $T$-dependence in the COM structure factor,
except for the very clear reduction of the compressibility on 
cooling (see below),
suggest the possibility that $T$-independent peaks in the
experimentally measured OTP structure factor are associated to 
COM features. Of course, the present model does not allow a 
straightforward comparison with experimental data (as performed for example in Ref.\cite{32mossa}), 
since proton and carbon atoms are not explicitly taken into account.

The large simulation box allows to precisely calculate the
$q\rightarrow0$  limit of the structure factor. As expected, we find that
$S_{_{COM}}(q\rightarrow0) = S_s(q\rightarrow0)/3$. The $T$-dependence of $S_s(q\rightarrow0)$ is
shown in the inset of Fig. \ref{fig:sq}.

\subsection{Self-Dynamics}
\label{sec:md:self}
\begin{figure}
\centerline{\psfig{figure=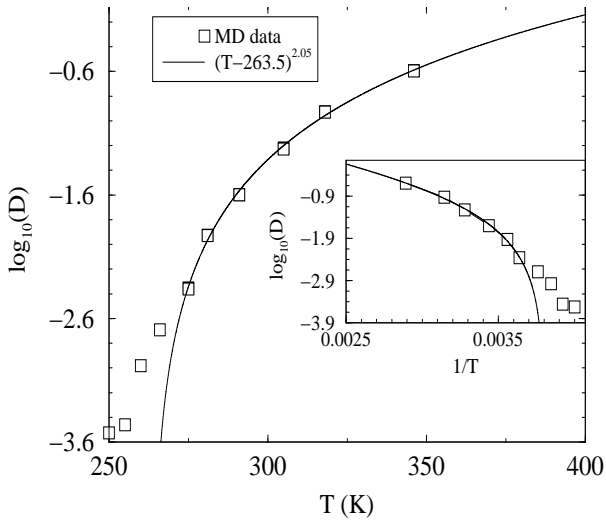,height=7.cm,width=8.cm,clip=,angle=0.}}
\caption{Temperature dependence of the diffusion coefficient, 
evaluated from the 
long time limit of the mean square displacement. The inset show the same data as a function of $1/T$.}
\label{fig:diff}
\end{figure}

Fig. \ref{fig:diff} shows the $T$ dependence of the diffusion
coefficient $D$, evaluated from the long time behavior of the mean
square displacement.  Compared to the LW original data, 
the range of $T$ where equilibrium states has been
simulated is larger and the precision in the determination of the
values of $D$ is improved.  The larger $T$-range shows that
two regions, which  differ in the $T$-dependence of $D$, are observed. 
Above $T=275$ $K$,
$D(T)$ is consistent with a power law behavior in $T-T_{MD}^c$, with
$T_{MD}^c=(265 \pm 1)K$ and exponent $\gamma=2.0 \pm 0.1$.  Below $T=275
K$, $D$ decreases on a slower pace, which could be approximated with
an Arrhenius law.  The two different behaviors reflect the dynamics
differences associated to the region of validity of MCT ($T>T_{MD}^c$)
and to the region ($T <T_{MD}^c$) where hopping phenomena become
relevant. The change in dynamical behavior shows up already for
$|T-T^c_{MD}|/T^c_{MD}= 0.06$, confirming the general expectation that
hopping phenomena may mask the critical dynamics close to the critical
MCT temperature.  As discussed at length in Ref.\cite{fspisa}, the
presence of hopping phenomena in this model makes impossible a
clear-cut identification of the critical dynamics, without making use
of information provided by the $T-$dependence of the
$\alpha$-relaxation phenomenon.  Similar changes in dynamical behavior
close to the critical MCT temperature have been recently reported in
models of silica\cite{kob-sio2} and water\cite{francis}.  It is worth
noting that in the original work of LW, hopping phenomena in the
orientational degrees of freedom were detected starting from $T=266$ 
$K$.

\begin{figure}
\centerline{\psfig{figure=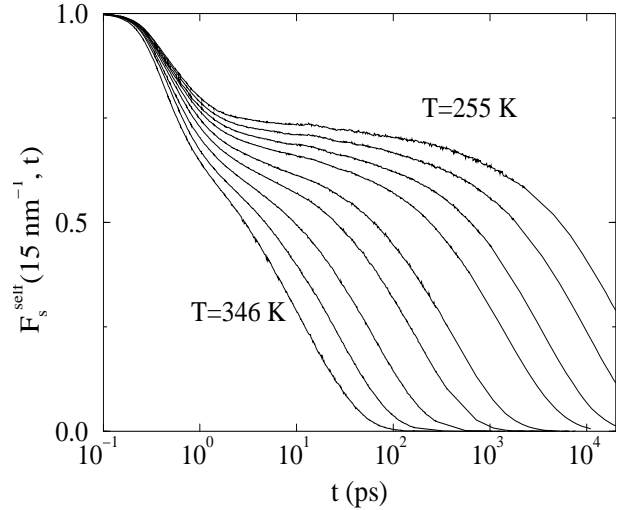,height=7.cm,width=8.cm,clip=,angle=0.}}
\caption{Site self-correlation function for all studied temperatures at
$q=15$ $nm^{-1}$ (the location of the first maximum of the site structure factor).}
\label{fig:self}
\end{figure}

The temperature dependence of the site self-correlation function
$F_s^{self}$ is
reported for a specific wavevector in Fig. \ref{fig:self}. Similar
figures characterize the time dependence of $F_s^{self}$  
at different values of $q$. 
The change in dynamics above and below $T_{MD}^c$ can be observed
also in
the so-called time-$T$ superposition graph, where curves for the same
correlator are shown as a function of a scaled time. Indeed, in the
$\alpha$-relaxation region the so-called time-temperature
superposition principle states that it is possible to represent the
$T$-dependence of an arbitrary
correlator $\phi(t)$ on a single master curve, by rescaling the time
via a single $T$-dependent timescale, i.e.

\begin{equation}
\phi(t)=\phi(t/\tau(T))
\label{eq:master}
\end{equation}

\begin{figure}
\centerline{\psfig{figure=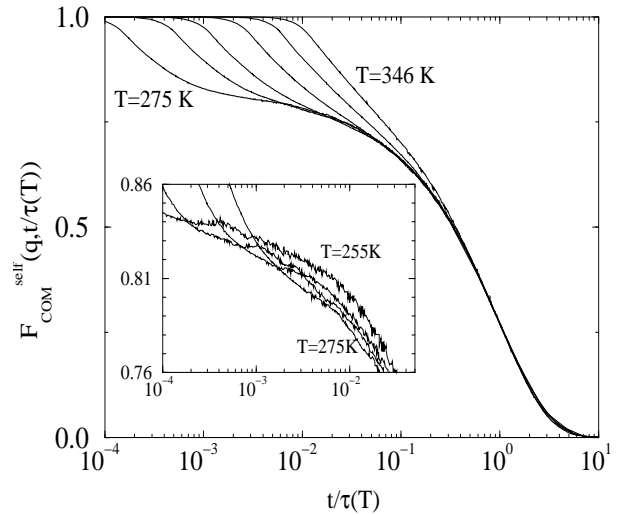,height=7.cm,width=8.cm,clip=,angle=0.}}
\caption{$\alpha$-scaling for the COM self-correlation function for different $T$ at $q=13~nm^{-1}$ --- the position of the first minimum in $S_{COM}(q)$.
The main plot shows, from rigth to left $T=346,318,305,291,281,275 K$.
The inset shows the region around the plateau for  $T=275, 266,260,255 K$.
}
\label{fig:alfaselfsite}
\end{figure}

\noindent
The scaling time $\tau(T)$ is 
conveniently defined\cite{KobAnd} as the time
when the correlation function has decayed to an arbitrary chosen
value, for example $1/e$.  Fig. \ref{fig:alfaselfsite} shows the
time-$T$ superposition for the $COM$ self density
correlation function  both above and below (inset) $T_{MCT}$.
Only for $T \ge 275$ $K$, where $D$ is well described by a
power-law, the time-$T$ superposition principle is well obeyed.
Below  $275$ $K$, deviations start to be noticeable and the
plateau value starts to increase.

To describe in a compact way the $q$ dependence of the
$\alpha$-relaxation and to make contact with the theoretical and
experimental evaluations, we present in the next two figures the
parameters which better describe the self density-density
correlation functions according to the Von Schweidler law and the
stretched exponential form. For a generic correlation function $\phi(t)$, 
the Von Schweidler law,
\begin{equation}
\phi(t) = f_{c} - h_{(1)} (t/\tau)^b + h_{(2)} (t/\tau)^{2b} + O(
(t/\tau)^{3b})
\label{eq:von}
\end{equation}

\noindent
describes, including the next to leading order corrections, the
departure from the plateau value $f_c$ in the early
$\alpha$-relaxation region.  The amplitudes $h_{(1)}$ and $h_{(2)}$
strongly depend on the physical features of the studied liquid and
have been explicitly calculated within MCT for several models (see for
example for hard spheres \cite{fuchs}).  The $\alpha$-relaxation time
scale $\tau$ is a temperature dependent parameter which, according to
MCT, scales as the inverse of diffusivity

\begin{equation}
\tau(T) \sim |T-T_c|^{-\gamma}
\label{eq:gamma}
\end{equation}

\noindent
The exponent $b$ is fixed by the value of $\gamma$\cite{review-glass}. 
The Kohlrausch-Williams-Watts stretched exponential form

\begin{equation}
\phi(t) = A_{_{K}} e^{-({t \over \tau_{_{K}}})^{\beta_{_{K}}}}
\label{eq:K}
\end{equation} 

\noindent
is often used to empirically fit the last part of the $\phi(t)$ decay.

Since all correlation functions obey the time-temperature
superposition principle, the parameters of the two functional forms
are calculated by fitting the $T=275$ $K $ correlation function in the
time interval $ 30 < t <  600~ps $ for the  Von Schweidler law  and 
in the time interval $t > 20$ $ps $ for the  stretched
exponential form. Similar results are obtained by fitting
the $T=266$ $K$ data.
The fit of the data according to Eq. \ref{eq:von} has been
performed by constraining the exponent $b$ to two different
values: the value $b=0.64$, 
obtained theoretically in Sec. \ref{sec:mct} for both 
approximations and the value $b=0.77$  
consistent with the exponent $\gamma$ extracted from the
analysis of the diffusivity data.
The fitting parameters $f_c$  does not depend
on the value of $b$. Except for a $q$-independent multiplicative factor,
also  the $q$-dependence of $h_1$ is independent of the choice of $b$. 

\begin{figure}
\centerline{\psfig{figure=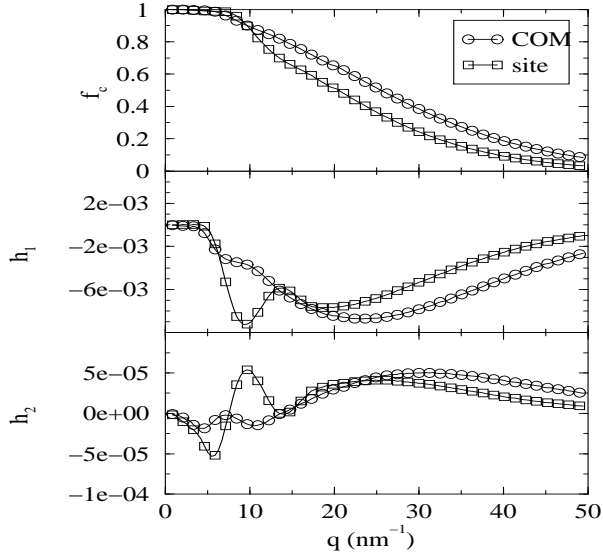,height=7.5cm,width=8.cm,clip=,angle=0.}}
\caption{Wavevector dependence of the fitting parameters according to
the Von Schweidler law  (Eq.~\protect\ref{eq:von}) for
the self correlation functions  $F_{s}^{self}(q,t)$ and  $F_{_{COM}}^{self}(q,t)$.~$T=275 K$.}
\label{fig:fitVSself}
\end{figure}

Fig. \ref{fig:fitVSself} shows the $q$-dependence of the fitting parameters
according to Eq. \ref{eq:von}. In the case of the self correlation function,
the non-ergodicity parameter $f_c(q)$ provides a description, in $q$-space, of
the confining cage. The width of the COM cage is larger than
the site one, supporting the view that the site experience an additional
delocalization associated to the hindered librational motions. Moreover, 
we note that 
the site non-ergodicity parameter is not 
described by a simple gaussian in $q$-space.
The $q$-dependence of $h_1$ and $h_2$ follows the oscillation of the
corresponding static structure factor.

\begin{figure}
\centerline{\psfig{figure=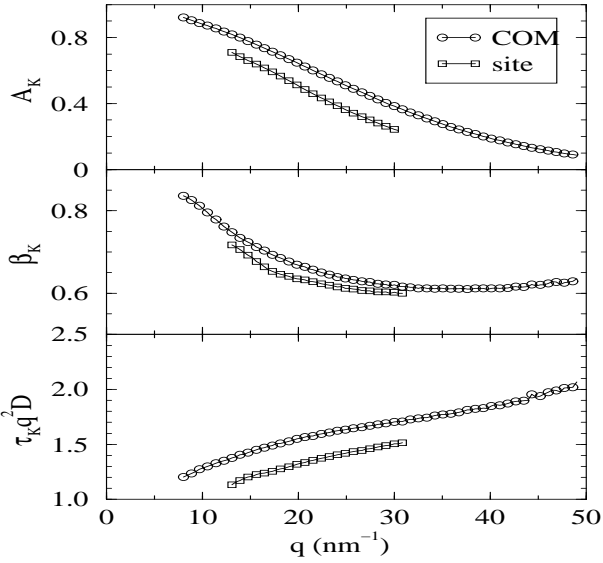,height=7.5cm,width=8.cm,clip=,angle=0.}}
\caption{Wavevector dependence of the fitting parameters
according to the stretches exponential form 
 (Eq.~\protect\ref{eq:K}) for$F_{s}^{self}(q,t)$ and  $F_{_{COM}}^{self}(q,t)$.$T=275 K$. }
\label{fig:fitSEself}
\end{figure}

Fig. \ref{fig:fitSEself} shows the
fitting parameters according to Eq. \ref{eq:K}. At small wavevectors the
extremely long decay of the correlation function, controlled by the diffusion
of the molecules over distances of the order of $q^{-1}$ does not allow 
an unbiased determination of the fitting parameters. From the wavevector range
where the fitting parameters are reliable, one can notice that, 
as commonly found, 
$A_k$ mimics $f_q$, that $\beta$ goes to $1$ at small 
$q$ and that the product $Dq^2\tau$ has a weak $q$-dependence.

\begin{figure}
\centerline{\psfig{figure=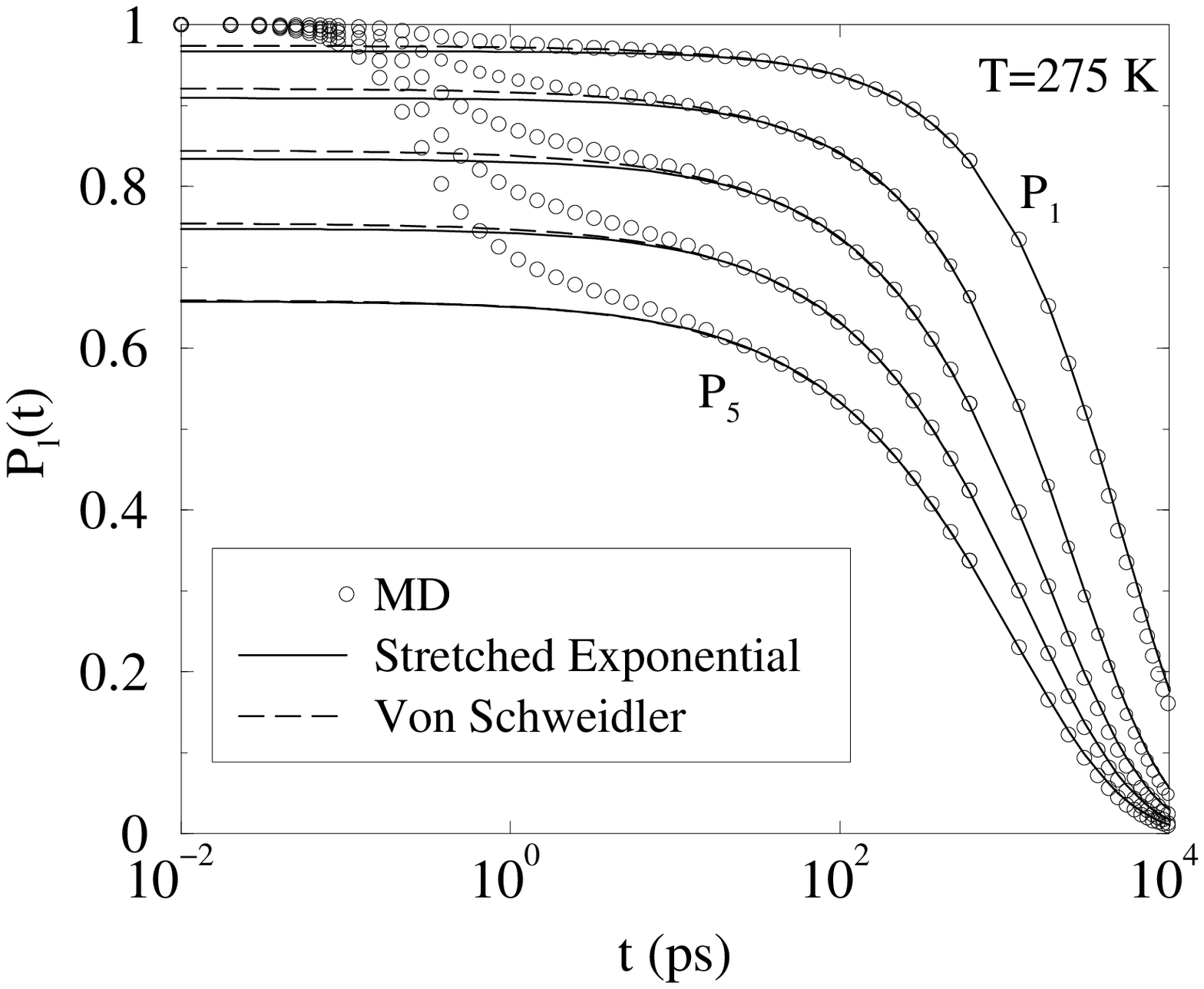,height=7.cm,width=8.cm,clip=,angle=0.}}
\centerline{\psfig{figure=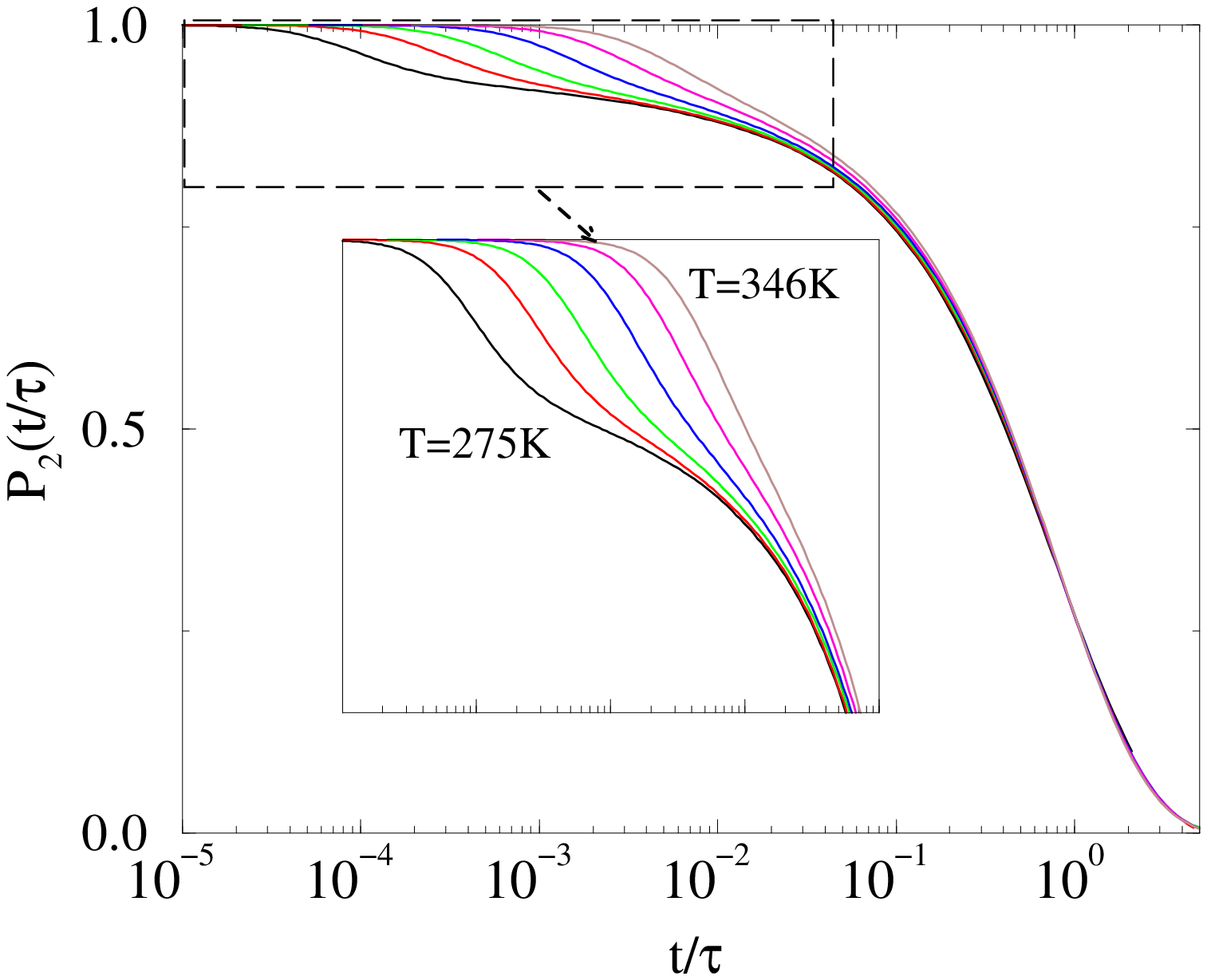,height=7.cm,width=8.cm,clip=,angle=0.}}
\caption{(top) Time dependence of the first five Legendre 
polynomials at $T=275 K$.(bottom) Time-Temperature superposition for the 
second-order Legendre polynomium.} 
\label{fig:p1p5}
\end{figure}

\begin{figure}
\centerline{\psfig{figure=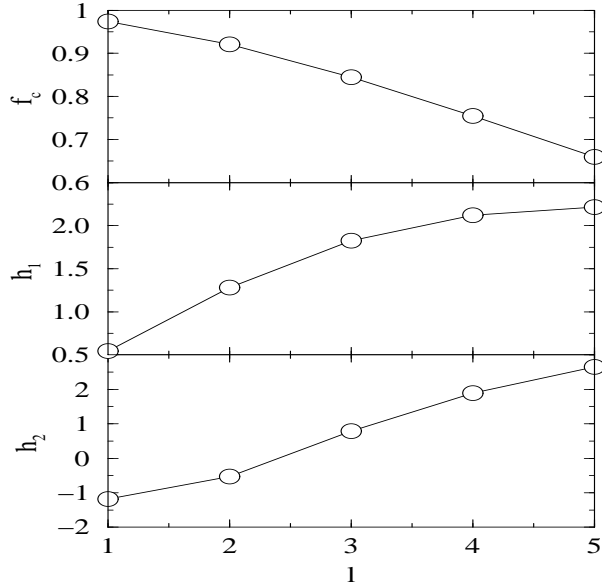,height=8.cm,width=8.cm,clip=,angle=0.}}
\caption{Fitting parameters according to the Von Schweidler law 
of the first five Legendre polynomials}
\label{fig:vslegendre}
\end{figure}

\begin{figure}
\centerline{\psfig{figure=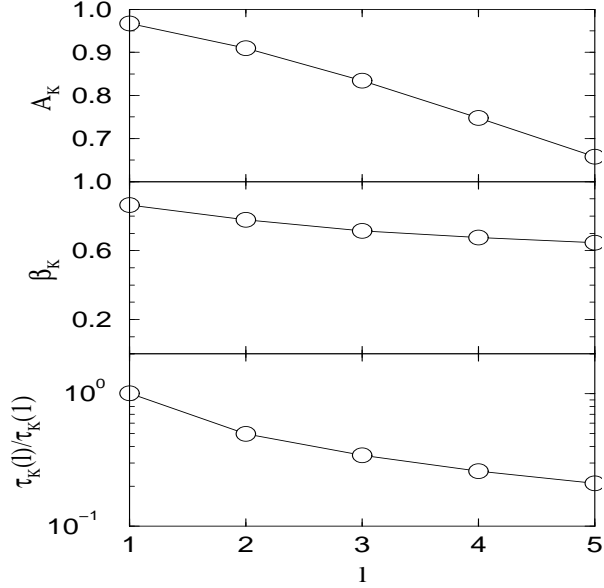,height=8.cm,width=8.cm,clip=,angle=0.}}
\caption{Legendre's order dependence of the fitting parameters according to
the stretches exponential form (Eq.~\protect\ref{eq:K})}
\label{fig:se-legendre}
\end{figure}

We also report the rotational dynamics of the LW molecule along the symmetry
axis. Fig. \ref{fig:p1p5}(top) shows the time dependence of 
the first five Legendre polynomials. We also show the corresponding fit
with the the stretched exponential and the Von Schweidler law.
For all angular correlators, the time-temperature superposition principle
holds beautifully, as shown in Fig.~\ref{fig:p1p5}(bottom).
Since the quality of the data is very good, even tiny discrepancies 
would be visible. For completeness, we report also the
fitting parameters according to the functional forms of 
Eqs.~\ref{eq:von}-\ref{eq:K} 
in Figs.~\ref{fig:vslegendre}-\ref{fig:se-legendre}.
The monotonic behavior of the fitting parameters in $l$ suggests that
for the LW potential, the scenario for the rotational dynamics is of the
strong hindrance type\cite{goetzeindrance}, as found also for the case
of SPC/E water\cite{noi-vigo,soper}.

The theoretical prediction of the wavevector dependence 
of the quantities reported in
Figs.~\ref{fig:fitVSself},\ref{fig:fitSEself},\ref{fig:vslegendre} and
\ref{fig:se-legendre} is one of the major challenge to 
the recent proposed molecular mode coupling 
theories\cite{schilling-vigo,site-MCT,franosch-dumb}.

\subsection{Collective Dynamics}

The collective density fluctuations are particularly relevant in the
description of liquids dynamics . The non-monotonic
wavevector dependence of the characteristic times provides indication of
the different timescale of the structural modes of the system
and their connection with the system characteristic lengths.

\begin{figure}
\centerline{\psfig{figure=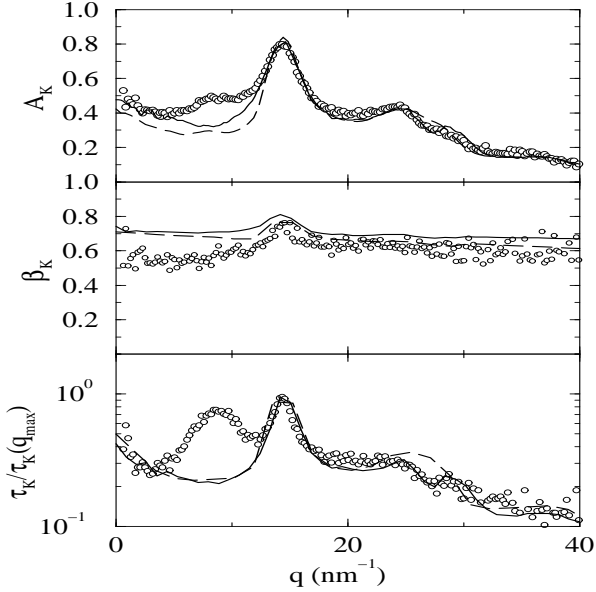,height=8.cm,width=8.cm,clip=,angle=0.}}
\caption{Wavevector dependence of the fitting parameters according to
the stretches exponential form (Eq.~\protect\ref{eq:K}) for
the normalized $F_{s}(q,t)$. Symbols are
MD results.
The full and dashed lines are
the prediction of the MCT using the site structure factor as input
and including or excluding the $c_3$ contribution.}
\label{fig:se-site-colle}
\end{figure}

We start discussing the site-site correlation functions, 
which can be compared
with the corresponding correlation functions calculated
according to MCT, as discussed in Sec.~\ref{sec:mct}.
In comparing with the  Von Schweidler and the stretched exponential
parameterizations,  the solutions
of the MCT equations (with and without the $c_3$ contribution), 
have been treated  similarly to the MD data.
Fig. \ref{fig:se-site-colle} shows the 
$q$-dependence of the stretched exponential parameters, $A_K$, $\beta_K$ and
$\tau_K$. We note that the theory captures the $q$-dependence of the 
$\alpha$-relaxation phenomenon close to the
maximum of the site-site structure factor. The large wavevector
region is also described in a reasonable way. The theory fails in describing
the non-ergodicity parameters in the region where the COM structure
factor has his peak position.  The slowest modes in the system are indeed
located around $9$ $nm^{-1}$ ($S_{_{COM}}$ peak position) and around $15$  
$nm^{-1}$ ($S_s$ peak position).  The role of the center of mass 
static correlation in controlling the
molecular dynamics has been observed in  recent experimental
studies on supercooled molecular liquids\cite{alba}.

The slowing down of the dynamics in the very low-$q$ region, below the
region of influence of the COM, is also predicted in a reasonable way 
by the site MCT developed in Sec.~\ref{sec:mct}.  
The stretching exponent parameter $\beta_K$ is the one which seems to
suffer more by the chosen approximations, which neglects the 
site-site intramolecular constrains. 
Indeed, the theoretical estimate is always higher than the MD results.
The prediction of a higher value of $\beta$ is consistent with the
reduced number of vertices included in the chosen MCT approach.
A more elaborate approach would require as input 
a large number of angular
static structure factor correlators\cite{mmct1} 
(which for the present molecule would make the theory difficult
to solve numerically) or the site-site partial static structure factors,
discriminating at least between the central and the two external sites.
The development of the molecular and site-site MCT approaches is taking place 
in the present days and
we look forward applying them to the LW model.

\begin{figure}
\centerline{\psfig{figure=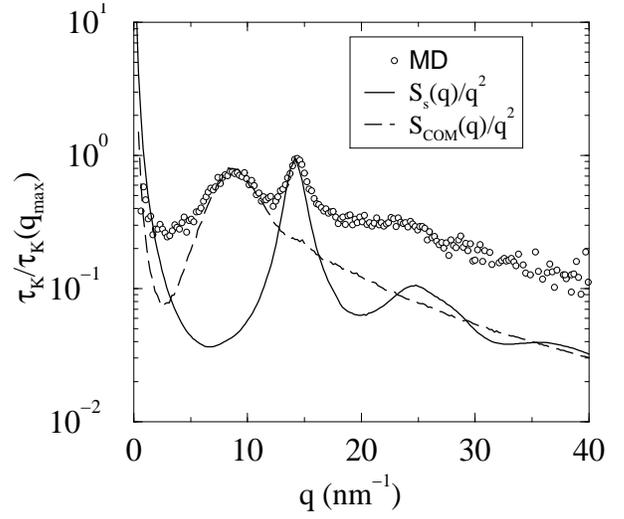,height=7.cm,width=8.cm,clip=,angle=0.}}
\caption{Comparison of the wavevector dependence of the $\alpha$-relaxation time with the de Gennes approximation for both site and COM.}
\label{fig:degennes}
\end{figure}

To substantiate the hypothesis that the disagreement between theory
and MD data in the region around $9$ $nm^{-1}$ arises from the 
approximation used, which neglects  
the constrains among sites due to the intramolecular bonding,
we compare in Fig. \ref{fig:degennes} the de Gennes predictions for
the COM and site correlation times with the corresponding MD data
(from Fig.~\ref{fig:se-site-colle}). It is clearly seen that the
dynamics in the wavevector region around $9$ $nm^{-1}$ is indeed
controlled by the COM static structure factor\cite{albadegennes}.

\begin{figure}
\centerline{\psfig{figure=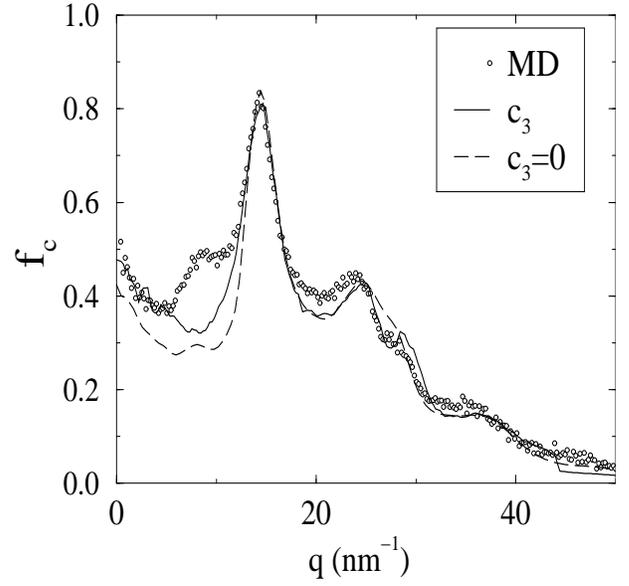,height=8cm,width=8.cm,clip=,angle=0.}}
\caption{Wavevector dependence of the 
non-ergodicity factor. The MD data has been calculated as
fitting parameters according to
the Von Schweidler law (\protect\ref{eq:von}) The full and dashed lines are
the prediction of the MCT using the site structure factor as input
and including or excluding the $c_3$ contribution.}
\label{fig:fqtheory}
\end{figure}

\begin{figure}
\centerline{\psfig{figure=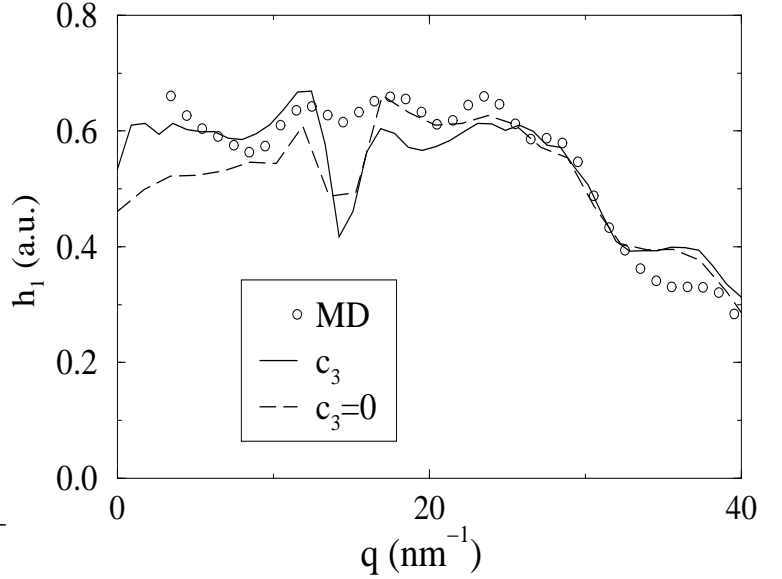,height=8cm,width=10.cm,clip=,angle=0.}}
\caption{Wavevector dependence of the amplitude $h_1$. The MD data (symbols) 
are the result of the fit 
according to the Von  Schweidler law (Eq.~\protect\ref{eq:von}). Lines are MCT predictions.}
\label{fig:h1theory}
\end{figure}

\begin{figure}
\centerline{\psfig{figure=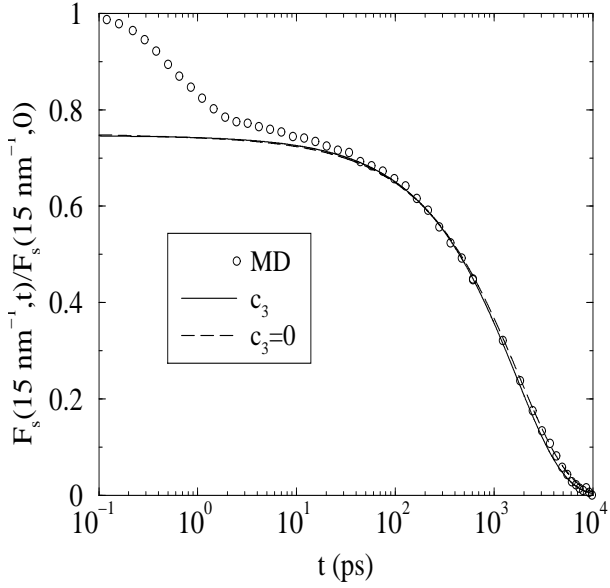,height=8cm,width=8.cm,clip=,angle=0.}}
\caption{Site-site 
$\alpha$-relaxation dynamics at the peak of the structure factor at 
$T=275 K$. Symbols are
MD results, lines are MCT predictions.}
\label{fig:dinamicamct}
\end{figure}

We next compare in Fig.~\ref{fig:fqtheory} and ~\ref{fig:h1theory} 
the prediction of MCT for the non ergodicity parameter and
for the amplitude of the Von Schweidler law with the MD results.
Again, we notice that the theoretical predictions 
satisfactorily describe the MD data except for the region where the
COM structure factor has its maximum. Including the $c_3$ contribution
improves the agreement in the hydrodynamic limit.
Around the peak position of $S_s(q)$, 
the agreement is very good, considering that neither
fitting nor scaling parameters are involved in the comparison.
In this wavevector region, the long-time dependence
of the collective site-site correlation function $F_s(q,t)$
is rather well predicted by the theory, as shown in 
Fig.~\ref{fig:dinamicamct}.

\begin{figure}
\centerline{\psfig{figure=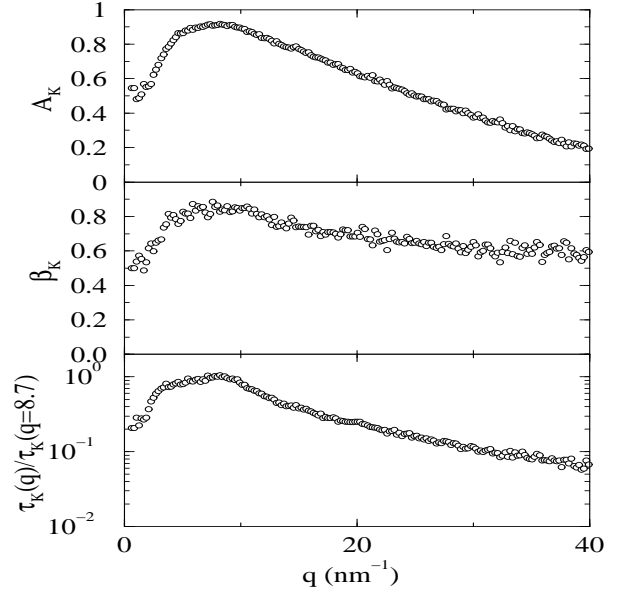,height=8cm,width=8.cm,clip=,angle=0.}}
\caption{Wavevector dependence of the fitting parameters according to
the stretches exponential form  (Eq.~\protect\ref{eq:K}) for
the normalized $F_{_{COM}}(q,t)$.}
\label{fig:se-cm-colle}
\end{figure}

\begin{figure}
\centerline{\psfig{figure=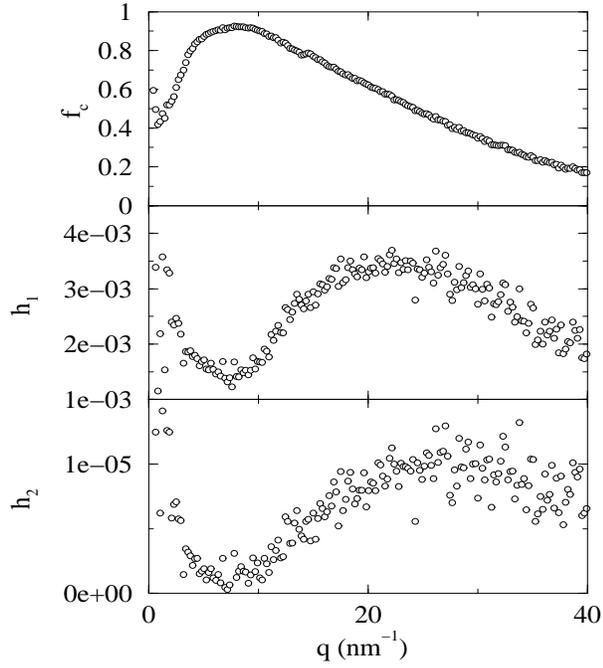,height=9cm,width=8.cm,clip=,angle=0.}}
\caption{Wavevector dependence of the fitting parameters according to 
Eq.~\protect\ref{eq:von} for the normalized $F_{_{COM}}(q,t)$.}
\label{fig:vs-cm-colle}
\end{figure}

We now turn to the COM correlators. These functions cannot be
compared with the theoretical predictions reported in Sec.\ref{sec:mct},
since this approximation neglects the molecular constrains. But
the $q$-dependence of the COM correlators is very 
relevant for comparisons with future molecular MCT predictions. 
Figs.~\ref{fig:se-cm-colle} and 
\ref{fig:vs-cm-colle} show the parameterization of the 
COM collective $\alpha$-relaxation using  Eqs. \ref{eq:K} and \ref{eq:von}.
The rather featureless shape of the COM structure, which as seen in Fig.
\ref{fig:sq} shows only a broad peak around $9$ $nm^{-1}$, carries on to the
the $q$-dependence of all fitting parameters.

\section{Conclusions}

The present study of the 
simple molecular model, introduced by Lewis and Wahnstr\"om 
to mimic the behavior of {\it ortho}-terphenyl,
provides a clean and simple reference model for
theoretical descriptions of the dynamics of molecular liquids in
supercooled states.

The simple MCT analysis reported in this article and 
compared in detail with the MD data shows that already the
MCT equations for atomic liquids are able to provide a 
rather accurate description of the site dynamics, except for 
the region around $9$ $nm^{-1}$ where the COM dynamics is dominant.
We have provided evidence that, in this region, the center of mass
fully controls the slow-dynamics in the liquid\cite{albadegennes}. 
Taking into account the triplet correlation function $c_3$ 
is shown to improve significantly only the small wavevector 
region. This is not surprising, since the 
usually employed approximation $c_3=0$ fails badly at small
wavevectors\cite{comment}. While this is not a major issue in the
intermediate wavevector range (i.e. for the caging physics) it
will become an important element in the attempt to 
extend (toward long wavelengths) the theoretical predictions.

The present theoretical work, which can be considered a zeroth 
order approximation to the molecular description, should be
followed by an accurate site-site MCT approach\cite{site-MCT} 
or by a full molecular MCT approach\cite{schilling}. 
Theoretical development 
along these two lines are taking place at a fast pace and
in the near future accurate theoretical predictions should be available
for the quantities studied in this article. For the present time,
we can only state that the characteristic MCT scenario appears  
able to describe the dynamics of this simple, yet complete, 
molecular model.

\section{Acknowledgment}
We acknowledge financial support from the INFM PAIS 98, PRA 99 and
{\it Iniziativa Calcolo Parallelo} and from MURST PRIN 98.

\noindent\begin{minipage}{8cm}
\begin{table}
\caption{For each temperature we list the diffusion coefficient, the density, the averages of the internal energy and of the external pressure and the lenght of the simulation time after the equilibration.}
\medskip
\begin{tabular}{|c||c|c|c|c|c|}
$T$  & $D$ & $\rho$ & U & P & $t_{sim}$ \\
$(K)$ &  $(10^{-5} cm^2/s)$ & $(g/cm^3)$ & $(kJ/mol)$ & $(MPa)$ & $(ns)$ \\
\tableline
255 & 0.00035 & 1.0851 & -77.6 & 0.02 & 41 \\
\tableline
260 & 0.0010 & 1.0822 & -77.3 & -1.2 & 52 \\
\tableline
266 & 0.0020 & 1.0790 & -77.0 & -1.0 & 37 \\
\tableline
275 & 0.0043 & 1.0763 & -76.5 & 5.4 & 37 \\
\tableline
281 & 0.012 & 1.0703 & -75.8 & 0.7 & 14 \\
\tableline
291 & 0.025 & 1.0650 & -75.2 & 2.0 & 16 \\
\tableline
305 & 0.059 & 1.0554 & -74.1 & 1.0 & 5 \\
\tableline
318 & 0.12 & 1.0439  & -72.9 & -2.3 & 7 \\
\tableline
346 & 0.25 & 1.0269 & -71.0 & -13.0 & 6
\end{tabular}
\label{table:table}
\end{table}
\end{minipage}


\begin{references}

\bibitem{KobAnd} W. Kob and H. C. Andersen,  Phys. Rev. E {\bf 51},
4626 (1995); {\it ibid.} Phys. Rev. E {\bf 52}, 4134 (1995).

\bibitem{Kob-LJ} M. Nauroth and W. Kob, Phys. Rev. E {\bf 55}, 675
(1997).

\bibitem{self} F. Sciortino, P. Gallo, P. Tartaglia, S.-H. Chen, Phys.
Rev. E {\bf 54}, 6331 (1996);  P. Gallo, F. Sciortino, P. Tartaglia, and
S.-H. Chen,  Phys. Rev. Lett. {\bf 76}, 2730 (1996).

\bibitem{collective} F. Sciortino, L. Fabbian, S.-H. Chen, and P. Tartaglia,
Phys. Rev. E {\bf 56}, 5397 (1997).

\bibitem{madden} M. C. Ribeiro and P. A. Madden,  J. Chem. Phys. {\bf 108}, 3256 (1998) and references therein.

\bibitem{heuer-doliwa} B. Doliwa and A. Heuer,  Phys. Rev. E {\bf 61}, 6898 (2000).


\bibitem{lai}
Lai S. K. Lai and G. F. Wang  Phys. Rev. E 58, 3072 (1998).
 
\bibitem{kob-dumb} S. K\"ammerer, W. Kob and R. Schilling,  Phys. Rev.
E {\bf 56}, 5397 (1997). 

\bibitem{heuer}
A. Heuer, Phys. Rev. Lett. {\bf 78}, 4051 (1997) ;
 S. Buechner and A. Heuer, Phys. Rev. E. {\bf 60}, 6507 (1999).


\bibitem{sri}
S. Sastri, P. Debenedetti and F. Stillinger, Nature, {\bf 393},
 554 (1998).

\bibitem{skt} F. Sciortino, W. Kob and P. Tartaglia,  Phys. Rev.  
Lett. {\bf 83 },  3214 (1999).

\bibitem{claudio} C. Donati, F. Sciortino, and P. Tartaglia,
 Phys. Rev. Lett. {\bf 85}, 1464 (2000). 


\bibitem{lanave} E. La Nave, A. Scala, F.W. Starr, F. Sciortino  and H.E.
Stanley, Phys. Rev. Lett. {\bf 84}, 4605 (2000).

\bibitem{selle} L. Angelani, R. Di Leonardo, G. Ruocco, A. Scala, F. Sciortino,
cond-mat/0007241; {\it ibid.} Phys. Rev. Lett. {\bf 85}, 5356 (2000).


\bibitem{review-glass} W. G\"{o}tze, in {\em Liquids, Freezing and Glass
Transition}, ed. J.P. Hansen, D. Levesque and J. Zinn-Justin, Les Houches
Session LI, 1989 (North-Holland, Amsterdam, 1991).

\bibitem{goetze-pisa} W.G\"{o}tze, J. Phys. Condens. Matter {\bf 11}, A1
(1999).


\bibitem{SIO2} F. Sciortino and W. Kob, 
cond-mat/0008024;  {\it ibid.} Phys. Rev. Lett. in press (2001).

\bibitem{ssm} L. Fabbian, F. Sciortino, F. Thiery and P. Tartaglia, 
Phys. Rev. E {\bf 57}, 1485 (1998).


\bibitem{mmct2} L. Fabbian, A. Latz, R. Schilling, F. Sciortino, P. Tartaglia,
and C. Theis,  Phys. Rev. E {\bf 60},  5768 (1999).

\bibitem{theis-new}  C. Theis, F. Sciortino, A. Latz, R. Schilling,
P. Tartaglia, cond-mat/0003508; {\it ibid.} Phys. Rev. E {\bf 62}, 1856 (2000)


\bibitem{winkler}  A. Winkler, A. Latz, R. Schilling, C. Theis,
cond-mat/0007276.

\bibitem{alder} B. J. Alder and  T. E. Wainwright,  J. Chem. Phys. {\bf 27},
1208 (1957).

\bibitem{lw}  L.J. Lewis and  G. Wahnstr\"om, Phys. Rev. E {\bf50},
3865 (1994).

\bibitem{plazek94} D. J. Plazek, C. A. Bero, I. C. Chay, J. Non-Cryst. Solids
 {\bf 172-174}, 181-190 (1994).

\bibitem{sillescu91} W. Petry, E. Bartsch, F. Fujara, M. Kiebel, H. Sillescu, 
and
B. Farago, Z. Phys. B {\bf 83}, 175-184 (1991).

\bibitem{sillescu93} J. Wuttke, M. Kiebel, E. Bartsch, 
F. Fujara, W. Petry, H. Sillescu,
Z. Phys. B {\bf 91}, 357-365 (1993).

\bibitem{sillescu95} E. Bartsch, F. Fujara, J. F. Legrand, 
W. Petry, H. Sillescu and J. Wuttke,
Phys. Rev. E {\bf 52}, 738 (1995).

\bibitem{toelle97} A. T\"olle, J. Wuttke, F. Fujara, and H. Schober, Phys. 
Rev. E {\bf 56}, 809 (1997).

\bibitem{masciovecchioPRL99} C. Masciovecchio, G. Monaco,
G. Ruocco, F. Sette, A. Cunsolo, M. Krisch,
A. Mermet, M, Soltwisch, and R. Verbeni, Phys. Rev. Lett.
 {\bf 80}, 544 (1998).


\bibitem{monacoPRL98} G. Monaco, C. Masciovecchio, G. Ruocco, and F. Sette, 
Phys. Rev. Lett. {\bf 80}, 2161 (1998). 

\bibitem{monacoPRL99} G. Monaco, D. Fioretto, C. Masciovecchio, G. Ruocco, and F. Sette, Phys. Rev. Lett.  
{\bf 82}, 1776 (1999).

\bibitem{site-MCT} S.-H. Chong and F.Hirata, Phys. Rev. E {\bf 58},
6188 (1998); Phys. Rev. E  {\bf 58}, 7296 (1998); 
Phys. Rev. E  {\bf 57}, 1691 (1998).

\bibitem{franosch-dumb} T. Franosch, M. Fuchs, W. G\"otze, M.R. Mayr and
A.P. Singh, Phys. Rev. E {\bf 56}, 5659 (1997).

\bibitem{schilling} R. Schilling and T. Scheidsteger, Phys. Rev. E
{\bf 56}, 2932 (1997); T. Scheidsteger and R. Schilling, Phil. Mag. B
{\bf 77}, 305 (1998);
C. Theis, Diploma Thesis, Johannes Gutenberg Universit\"atMainz (1997).



\bibitem{walstrom2}  
G. Wahnstr\"om and L. J. Lewis,
Prog. Theor. Phys. Suppl. {\bf 126}, 261 (1997).

\bibitem{barrat-c3} 
J.-L. Barrat, W. G\"otze, and A. Latz,
J. Phys.: Condens. Matter {\bf 1}, 7163 (1989).


\bibitem{hansen} J.P. Hansen and I.R. McDonald, {\em Theory of Simple
Liquids} (Academic Press, London, 1986), 2nd Edition.

\bibitem{numerics} A. P. Singh, {\em Die St\"{o}rung der $\beta$-Dynamik
durch oszillierende Moden}, Diploma thesis, TU M\"{u}nchen, 1995.


\bibitem{32mossa} S. Mossa, R. Di Leonardo, G. Ruocco, M. Sampoli,
Phys. Rev. E {\bf 62}, 612 (2000). S. Mossa et al (preprint).

                                                                      
\bibitem{fspisa} 
F. Sciortino and P. Tartaglia,
J. Phys.: Condens. Matter {\bf 11}, A261 (1999).

\bibitem{kob-sio2} J. Horbach and W. Kob, 
Phys. Rev. B {\bf 60}, 3169 (1999).


\bibitem{francis} F. W. Starr, S. Harrington, F. Sciortino, and H. E.  Stanley,
Phys. Rev. Lett. {\bf 82}, 3629 (1999). 

\bibitem{fuchs} M. Fuchs, I. Hofacker and A. Latz,  Phys. Rev. A {\bf
45}, 898 (1992).

\bibitem{goetzeindrance} 
W. Gotze, A. P. Singh, Th. Voigtmann Phys. Rev. E 61, 6934 (2000).


\bibitem{noi-vigo} L. Fabbian, F. Sciortino and P. Tartaglia, 
J. Non-Cryst. Solids {\bf 235-237}, 350 (1998).

\bibitem{soper} F. Sciortino,
Chem. Phys. {\bf 258}, 295-302 (2000).


\bibitem{schilling-vigo} C. Theis and R. Schilling, J. Non-Cryst.
Solids{\bf 235-237}, 106 (1998).

\bibitem{alba}
D. Morineau, C. Alba-Simionesco, J. Chem. Phys. {\bf 109}, 8494 (1998);
D. Morineau,   C. Alba-Simionesco, M.C. Bellissent-Funel and M.F. Lautie,
Europhys. Lett. {\bf 43}, 195 (1998).


\bibitem{mmct1} L. Fabbian, A. Latz, R. Schilling, F. Sciortino, P. Tartaglia,
and C. Theis, Phys. Rev. E {\bf 62}, 2388 (2000). 



\bibitem{albadegennes} C. Alba-Simionesco, A. T\"olle,  D. Morineau,
B. Farago and G. Coddens, {\it ``de Gennes'' narrowing in supercooled molecular liquids: Evidence for center-of-mass dominated slow dynamics} preprint (2000).


\bibitem{comment} The wavevector dependence of the triplet density fluctuations
$\langle \rho_s({\bf k}) \rho_s({\bf p}) \rho_s({\bf q}) \rangle $
fully determines the wavevector dependence of the
static structure factor (i.e. the pair density fluctuations) via the
Born-Green-Yvon (BGY) equation \protect\cite{tosi}.
It is easy to show that 
the approximation $c_3=0$, when used in conjunction with 
the BGY equation, produces an implicit equation for
the static structure factor which does not depend on density.
The approximation $c_3=0$ should not be used in the hydrodynamic limit.

\bibitem{tosi} A. N.H.March and M.P.Tosi, Atomic dynamics in liquids, p.23
Dover (NY) 1991. 



\end{references}
\end{document}